\documentclass[aps,prl,reprint,groupedaddress]{revtex4-1}

\usepackage{color}
\usepackage{graphicx}

\begin{document}


\title{Computation of anomalous scaling exponents of turbulence \\ from self-similar instanton dynamics}


\author{Alexei A. Mailybaev}
\affiliation{%
Instituto Nacional de Matem\'atica Pura e Aplicada -- IMPA, Rio de Janeiro,
Brazil
\footnote{Estrada Dona Castorina 110, 22460-320 Rio de Janeiro, RJ, Brazil. 
Phone:~+55\,21\,2529\,5070, Fax:~+55\,21\,2529\,5075, E-mail: alexei@impa.br}
}%
\affiliation{%
Institute of Mechanics, Lomonosov Moscow State University, Russia
}%


\date[]{E-mail: alexei@impa.br,\ \ \ \today}

\begin{abstract}
We show that multiscaling properties of developed turbulence in shell models, 
which lead to anomalous scaling exponents in the inertial range, 
are determined exclusively by instanton dynamics. 
Instantons represent correlated extreme events localized in space-time, whose 
structure is described by self-similar statistics with a single universal scaling exponent. 
We show that anomalous scaling exponents appear due to the process of instanton creation. 
A simplified model of instanton creation is suggested, 
which adequately describes this anomaly. 
\end{abstract}


\maketitle

One of the major open problems in hydrodynamic turbulence is the explanation of small-scale 
intermittency characterized by scaling exponents of structure functions~\cite{frisch1995turbulence}. 
Deviation of the scaling exponents from Kolmogorov's theory of 1941 based on dimensional 
considerations is called the anomaly. The mechanism responsible for this anomaly is still 
not well understood despite of strong theoretical effort, 
e.g., \cite{smith1998renormalization,*l1998computing,*l2000analytic,*adzhemyan2005improved}. 
Anomalous scaling is explained for a class of linear problems including the 
case of Kraichnan passive scalars using the notion of zero modes~\cite{falkovich2001particles}. 
However, there are essential theoretical difficulties in application of this method to the nonlinear Navier-Stokes turbulence~\cite{benzi2003intermittency,*angheluta2006anomalous}.

Simplified models may help to reveal the main physical mechanism leading to intermittency and its relation to nonlinear structure of the system. In this respect, shell models~\cite{biferale2003shell,gledzer1973system,*ohkitani1989temporal,l1998improved} are very successful to reproduce statistical properties, which agree qualitatively and quantitatively with laboratory experiments for homogeneous isotropic turbulence. Also, shell models allow for reliable numerical simulation at very high Reynolds numbers. 

In this paper, we show that the anomalous scaling exponents in shell models of turbulence can be defined in terms of velocity maximums, which split into subgroups with self-similar statistics. This splitting allows to recognize instantons as ``elementary particles'' of the turbulent statistics. Instanons are correlated extreme events localized in space-time and describing a blowup in inviscid limit~\cite{siggia1978model,*nakano1988,*dombre1998intermittency,*l2001outliers,*mailybaev2012}. Instanton structure is described by a single universal scaling exponent. We establish the exact relation between the scaling anomaly and the process of instanton creation and derive a general formula for scaling exponents for a simplified model of instanton creation. 

We focus on the Sabra shell model of turbulence~\cite{l1998improved}
\begin{equation}
\begin{array}{rl}
\displaystyle
\frac{du_n}{dt} 
\displaystyle
= & i(ak_{n+1}u_{n+2}u_{n+1}^*+bk_{n}u_{n+1}u_{n-1}^*
\\[5pt] &\displaystyle
-ck_{n-1}u_{n-1}u_{n-2})-\nu k_n^2u_n+f_n.
\end{array}
\label{eq1}
\end{equation} 
Here $u_n$ is the complex shell velocity, which can be understood as the Fourier component 
of the velocity field at the shell wavenumber $k_n = k_0\lambda^n$, $f_n$ is 
the forcing term restricted to the first shells, $\nu$ is the viscosity, 
and  $n = 1,\ldots,n_{\max}$ with large $n_{\max}$. 
Traditional choice of parameters is $\lambda = 2$, $a = 1$, $b = c = -0.5$, 
in which case the ``energy'' $\sum_n|u_n|^2$ and ``helicity'' $\sum_n (-)^nk_n|u_n|^2$ are conserved 
in the inviscid system with no forcing. 

\begin{figure}
\centering\includegraphics[width=0.48\textwidth]{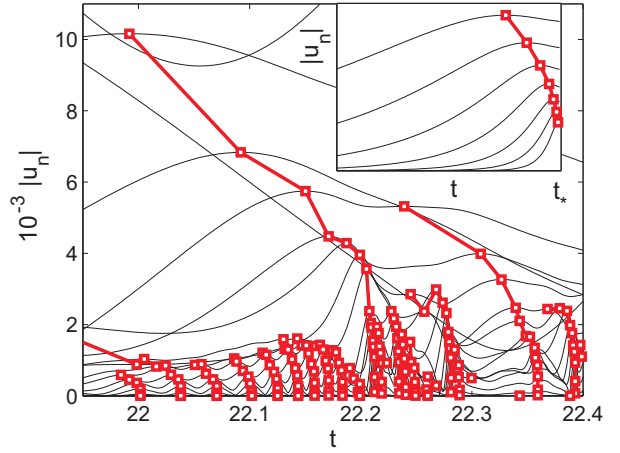}
\caption{\label{fig1} (Color online) Typical turbulent dynamics of the shell model in the inertial range. 
Red squares mark maximums of speed amplitudes. 
Instantons are identified by sequences of maximums following in increasing time and shell number.
Inset shows the universal asymptotic form of the blowup propagating into unperturbed 
medium (ideal instanton) for inviscid system.}
\end{figure}

The inviscid shell model possesses the self-similar solution determined up to 
phase symmetry by~\cite{siggia1978model,*nakano1988,*dombre1998intermittency,*l2001outliers,*mailybaev2012}
\begin{equation}
u_n(t) = -iu_*k_n^{-y_0}f(u_*(t_*-t)k_n^{1-y_0}),\quad t \le t_*,
\label{eq3}
\end{equation}
where the parameters $u_* > 0$, $t_*$ and the function $f(u)$ are real and $y_0 = 0.281$, see Fig.~\ref{fig1}(inset).  
Solution (\ref{eq3}) represents the universal asymptotic form of the blowup for the 
inviscid shell model~\cite{dombre1998intermittency,mailybaev2012}. 
It describes a stable front reaching the smallest scale in finite time 
and leaving behind the trail $u_n \propto k_n^{-y_0}$. 
Note that no singularity appears at $t = t_*$ for a particular shell number $n$, 
since all speeds $u_n(t_*)$ are finite. 

\begin{figure}
\centering\includegraphics[width=0.47\textwidth]{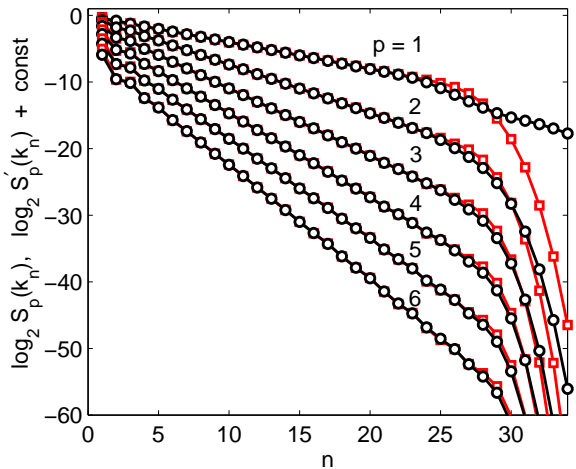}
\caption{\label{fig2} (Color online) Structure functions determined 
in terms of shell speeds and speed maximums. 
Shown are the plots of $\log_2 S_p(k_n)$ (red squares) and $\log_2 S'_{p}(k_n)$ (black circles) 
for $p = 1,\ldots,6$. 
The plots are shifted vertically to demonstrate equivalence of scaling 
laws in the inertial range.}
\end{figure}

Fig.~\ref{fig1} shows shell speed amplitudes in turbulent regime with a high Reynolds number. 
In numerical simulation, we used $n_{\max} = 34$ shells with $k_0 = 2^{-6}$, the forcing $f_1 = 2f_2 = (5+5i)10^{-3}$, the viscosity $\nu = 5\times 10^{-10}$ and the time interval $T = 90000$. 
Numerical data displays the well-known multiscaling properties characterized by the structure functions 
\begin{equation}
S_p(k_n) 
\equiv \lim_{T\to\infty}\frac{1}{T}\int_0^T|u_n|^pdt
\propto k_n^{-\zeta_p} 
\propto \lambda^{-n\zeta_p},
\label{eq2}
\end{equation}
where the scaling exponents $\zeta_p$ exhibit nonlinear dependence on $p$, Fig.~\ref{fig2}. 
Expression (\ref{eq2}) is valid in the inertial range $n_f < n < n_v$, 
where $n_f \sim 4$ is the largest number of the shell affected by forcing (forcing range $n \le n_f$) 
and $n_v \sim 25$ is the smallest shell number affected by viscosity (viscous range $n \ge n_v$). 

Turbulent dynamics represents a series of bursts, which differ in intensity and duration 
leading to the characteristic intermittent behavior. 
It is argued~\cite{gilson1997towards,*daumont2000instanton} that these bursts 
(called \textit{instantons}) may have self-similar structure described 
statistically by expression (\ref{eq3}), where the shape $f(u)$ and 
scaling exponent $y_0$ are affected by random velocity background. 
For numerical study of instanton statistics, 
we suggest to identify each instanton with a sequence of local maximums $v_n = \max |u_n(t)|$ 
at times $t_n$ following in increasing order $t_{n_0} \le t_{n_0+1} \le \cdots \le t_{n_1}$, Fig.~\ref{fig1}. No maximums of $|u_{n}(t)|$ and $|u_{n+1}(t)|$ are allowed in the interval $t_{n} < t < t_{n+1}$. Each instanton is created at some shell number $n_0$ and either reaches the viscous range or annihilates at a shell number $n_1$ in the inertial range. Using this rule, we group all maximums of velocity amplitudes into instantons indexed by $\alpha$. 

Contribution of self-similar solution (\ref{eq3}) to the structure function integral 
in Eq.~(\ref{eq2}) is  
\begin{equation}
\int_{-\infty}^{t_*}|u_n|^pdt \propto (u_*k_n^{-y_0})^{p-1} k_n^{-1} \propto v_n^{p-1}k_n^{-1}, 
\label{eq4}
\end{equation}
where $v_n = \max|u_n(t)|\propto u_*k_n^{-y_0}$. 
The scaling law in Eq.~(\ref{eq4}) is independent of the function $f(u)$ and exponent $y_0$. 
Assuming self-similar structure of instantons, 
we can use Eq.~(\ref{eq4}) to compute structure functions in Eq.~(\ref{eq2}) as
\begin{equation}
S_p(k_n) \propto \lim_{T \to \infty}\frac{1}{Tk_n}\sum_\alpha v_n^{p-1}, 
\label{eq6}
\end{equation}
where the sum is taken over all local maximums (all instantons $\alpha$).
This suggests the alternative definition of structure functions in terms of velocity maximums as
\begin{equation}
S'_p(k_n) \equiv \lim_{T\to \infty}\frac{1}{Tk_n}\sum_\alpha v_n^{p-1} \propto k_n^{-\zeta_p} \propto \lambda^{-n\zeta_p}, 
\label{eq7}
\end{equation}
with the same scaling exponents $\zeta_p$.
Scaling law (\ref{eq7}) is in full agreement with numerical results, Fig.~\ref{fig2}. 
If the sum in Eq.~(\ref{eq7}) is taken over all \textit{stable} instantons, which propagate from the creation shell number $n_0$ all the way to the viscous range, 
the scaling law (\ref{eq7}) remains valid with almost the same exponents $\zeta_p$ 
(decreased approximately by 0.02). Thus, instantons annihilating in the inertial range have very small effect on the scaling law in Eq.~(\ref{eq7}) and will not be included in the sum $\sum_\alpha$ from now on.  
These instantons will be discussed later.

\begin{figure}
\centering\includegraphics[width=0.48\textwidth]{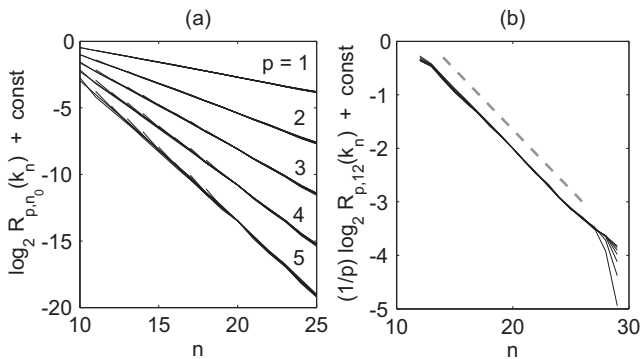}
\caption{\label{fig3} 
Single-scale statistics of instantons created at the same shell.
(a) Plots of $\log_2 R_{p,n_0}(k_n)$ for structure functions  corresponding to instantons created at the shell $n_0$. 
Shown are the plots for $8 \le n_0 \le 17$, which collapse after 
vertical translation to a straight line 
for each $p = 1,\ldots,5$. 
(b)  Plots of $(1/p)\log_2 R_{p,12}(k_n)$ for $p = 1,\ldots,7$, which collapse after vertical translation to a straight line 
of slope $-y$.}
\end{figure}

Statistical properties of the instantons can be understood by considering the structure functions
\begin{equation}
R_{p,n_0}(k_n) \equiv \lim_{T \to \infty}
\frac{1}{T}\sum_{\alpha(n_0)} v_n^{p}, 
\quad n \ge n_0,
\label{eq9}
\end{equation}
where the sum is restricted to stable instantons created in the shell $n_0$. 
Log-log plots of the functions $R_{p,n_0}(k_n)$ are presented in Fig.~\ref{fig3}a. 
After the vertical translation, the curves with equal $p$ and different $n_0$  collapse to the same straight line. 
The slope of this line is proportional to $p$, see Fig.~\ref{fig3}b, 
implying the power law dependence  
\begin{equation}
R_{p,n_0}(k_n) = c_{p,n_0}\lambda^{(n_0-n)py},\quad
y = 0.225,
\label{eq10}
\end{equation}
in inertial range.
Linear dependence of the scaling exponent on $p$ means that multiscaling of structure 
functions in Eqs.~(\ref{eq2}) and (\ref{eq7}) is not a stastistical property for a 
particular instanton. On the contrary, statistics of instantons 
created in the same shell 
is self-similar 
with a single scaling exponent $y$. Note that $y$ is close but not equal to 
the scaling exponent $y_0 = 0.281$ of the instanton in Eq.~(\ref{eq3}) propagating 
in undisturbed velocity background.

\begin{figure}
\centering\includegraphics[width=0.47\textwidth]{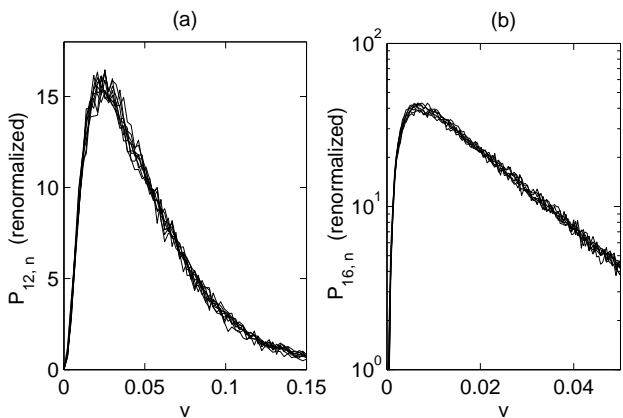}
\caption{\label{fig4} Self-similarity of PDFs describing local maximums 
for instantons created at shell $n_0$. 
(a) Renormalized PDFs $\lambda^{(n_0-n)y}P_{n_0,n}(\lambda^{(n_0-n)y}v)$ computed  
for $n_0 = 12$ and $n = 12,\ldots,20$ collapse to a single curve. 
(b) Similar plots for $n_0 = 16$ and $n = 16,\ldots,22$ presented in log-scale 
to show exponential form and good match of the tails.}
\end{figure}

Self-similarity of instantons can be confirmed by considering the probability distribution functions (PDFs) determining the probability $P_{n_0,n}(v)dv$ to sample a local maximum $v = \max|u_n(t)|$ belonging to the instanton created in the shell $n_0$. 
Self-similarity for PDFs implies
\begin{equation}
P_{n_0,n}(v) = \lambda^{(n-n_0)y}P_{n_0,n_0}(\lambda^{(n-n_0)y} v),
\quad n > n_0,
\label{eq10b}
\end{equation}
which is in full agreement with numerical results, Fig.~\ref{fig4}.

By definitions in Eqs.~(\ref{eq7}) and (\ref{eq9}),
\begin{equation}
\begin{array}{rl}
S'_{p}(k_n) & 
\displaystyle
= k_n^{-1}\sum_{n_0 = 1}^n R_{p-1,n_0}(k_n) 
\\&
\displaystyle
= k_n^{-1}\sum_{n_0 = 1}^n c_{p-1,n_0}\lambda^{(n_0-n)(p-1)y},
\end{array}
\label{eq11}
\end{equation}
where we also used Eq.~(\ref{eq10}). The coefficients $c_{p,n_0}$ are expressed from 
Eq.~(\ref{eq11}) as
\begin{equation}
c_{p,n_0} 
= k_{n_0}S'_{p+1}(k_{n_0})-\lambda^{-py}k_{n_0-1}S'_{p+1}(k_{n_0-1}).
\label{eq13}
\end{equation}
Using the scaling laws (\ref{eq7}) in Eq.~(\ref{eq13}) yields 
\begin{equation}
c_{p,n_0} \propto \lambda^{n_0(1-\zeta_{p+1})}
\label{eq12}
\end{equation}
for fixed $p$, 
which is confirmed numerically.
We see that, due to self-similar structure of instantons, multiscaling is attributed to the coefficients $c_{p,n_0}$ in Eq.~(\ref{eq10}) depending on creation 
shell number $n_0$. 

We showed that scaling properties of turbulence 
in the inertial range are determined exclusively by 
statistics of instantons. 
In this picture, instantons play the role of self-similar ``elementary particles'' 
of turbulent dynamics, and the anomaly must arise in the process of instanton creation. 
According to Eqs.~(\ref{eq9})--(\ref{eq10b}), characteristic amplitude 
of the instanton is given by $u_n \sim v^*\lambda^{(n_0-n)y}$ 
with the universal exponent $y = 0.225$, where $n_0$ is the creation shell number and $v_*$ 
is the initial amplitude. From Eq.~(\ref{eq4}), we find the characteristic lifetime of the 
instanton in shell $n$ as $v_n^{-1}k_n^{-1}$. 
Then the total time occupied by instantons in the shell number $n$ 
is found as
\begin{equation}
\sum_{\alpha} v_n^{-1}k_n^{-1} = TS'_0(k_n) 
\propto Tk_n^{-\zeta_0} = T,
\label{eq15}
\end{equation}
where we used Eq.~(\ref{eq7}) with $\zeta_0 = 0$. Relation (\ref{eq15}) was also confirmed numerically. 
It shows that instantons are dense in the phase space $(n,t)$ 
and suggests considering dynamics in the inertial range as a ``gas'' of interacting instantons. 
Note that instanton amplitude is inverse proportional to its lifetime. Therefore, the observed 
dynamics with long periods of low activity (intermittency) 
is due to weak instantons with long lifetimes. 

Understanding of instanton creation process is a complicated problem. As we already mentioned, 
unstable instantons, which annihilate in the inertial range, have very small influence on scaling exponents. However, numerical simulation shows that the number of maximums corresponding to unstable instantons is not negligible: it decreases with $n$ from $30$ to $5\%$ of the total number.  Unstable instantons, which survive few shells, can be viewed as random fluctuations. 
Unimportance of these fluctuations for scaling laws 
suggests that statistics is governed primarily by instanton interactions rather 
than by random background as suggested in~\cite{gilson1997towards,*daumont2000instanton}.

Below we propose a very simple model of instanton creation leading to multiscaling. 
Let us consider the PDF $P_n(v)$ of all instantons in shell $n$ given by 
\begin{equation}
N_nP_n(v) = 
\sum_{n_0 = 1}^{n}M_{n_0}P_{n_0,n}(v),
\label{eq17}
\end{equation}
where $N_n$ is the total number of instantons (maximums) in shell $n$ 
and $M_{n_0} = N_{n_0}-N_{n_0-1}$ is the number of instantons created in shell $n_0$. Our model is based on the assumption that instantons are created in self-consistent way determined by their distribution in the same shell,  
\begin{equation}
P_{n,n}(v) = \lambda^{x}P_{n}(\lambda^{x}v).
\label{eq16}
\end{equation}
This assumption has satisfactory agreement with the numerical results for 
the scaling exponent $x = 0.21$.

Now we can find the structure functions.  
Using Eqs.~(\ref{eq16}), (\ref{eq10b}) in the right-hand side of Eq.~(\ref{eq17}), we write
\begin{equation}
\begin{array}{rl}
N_nP_n(v) 
&\displaystyle 
\!\!= 
\lambda^{x}M_nP_{n}(\lambda^{x}v)+\lambda^{y}\sum_{n_0 = 1}^{n-1} M_{n_0}P_{n_0,n-1}(\lambda^{y}v) 
\\[12pt]
& \!\!= \lambda^{x}M_nP_{n}(\lambda^{x}v)+\lambda^{y}N_{n-1}P_{n-1}(\lambda^{y}v),
\end{array}
\label{eq17b}
\end{equation}
where we used Eq.~(\ref{eq17}) again in the last equality. 
Structure functions (\ref{eq7}) can be expressed as
\begin{equation}
S'_{p}(k_n) = \lim_{T \to \infty}\frac{N_n}{Tk_n}\int v^{p-1} P_{n}(v)dv.
\label{eq18}
\end{equation}
Multiplying both sides of Eq.~(\ref{eq17b}) by $(Tk_n)^{-1}v^{p-1}dv$ and integrating yields
\begin{equation}
S'_p(k_n) 
= \lambda^{x(1-p)}
\frac{M_n}{N_n}
S'_p(k_n)
+\lambda^{y(1-p)-1}S'_p(k_{n-1}).
\label{eq20}
\end{equation}
Using the exact result $\zeta_3 = 1$ following from the constant energy 
flux condition~\cite{l1998improved}, we substitute $p = 3$ and $S'_3(k_n) \propto \lambda^{-n}$ in Eq.~(\ref{eq20}) and obtain  
\begin{equation}
M_n/N_n = \lambda^{2x}(1-\lambda^{-2y}).
\label{eq21}
\end{equation}
Substituting Eq.~(\ref{eq21}) into Eq.~(\ref{eq20}), we obtain  
\begin{equation}
S'_p(k_n) 
= \frac{\lambda^{y(1-p)-1}}{1-\lambda^{x(3-p)}(1-\lambda^{-2y})}S'_p(k_{n-1}).
\label{eq22}
\end{equation}
This relation implies the scaling law in Eq.~(\ref{eq7}) with
\begin{equation}
\zeta_p 
= 1+y(p-1)+\log_\lambda(1-\lambda^{x(3-p)}(1-\lambda^{-2y})).
\label{eq23}
\end{equation}
One can see that $\zeta_3 = 1$. The condition $\zeta_0 = 0$ requires
\begin{equation}
x 
= (\log_\lambda(1-\lambda^{y-1})-\log_\lambda(1-\lambda^{-2y}))/3 = 0.211,
\label{eq24}
\end{equation}
which is the value we used above.

\begin{figure}
\centering\includegraphics[width=0.45\textwidth]{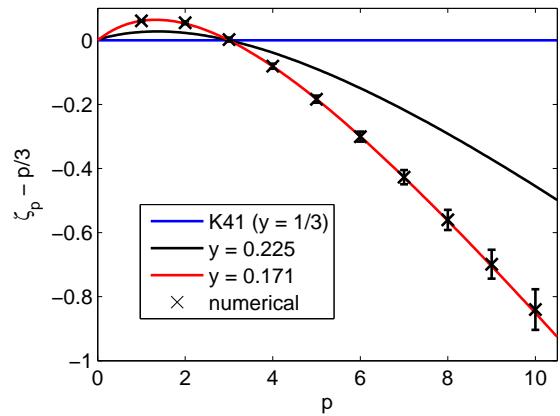}
\caption{\label{fig7} (Color online) Anomalous part $\zeta_p-p/3$ of scaling exponents.  
Solid lines show the result of the simplified theory of instanton creation with different 
values of parameter $y$. Error bars represent the result of numerical simulation.}
\end{figure}

Expression (\ref{eq23}) describes anomalous (nonlinear) dependence of scaling exponents on $p$ in qualitative agreement with the numerical data, Fig.~\ref{fig7}.  
The theory of instanton creation based on relation (\ref{eq16}) is approximate and does 
not take into account, e.g., dependence on the PDFs in neighboring shells. 
Thus, we did not expect very good quantitative agreement. 
Scaling exponents given by formulas (\ref{eq23}), (\ref{eq24}) depend on a single parameter $y$. 
It is interesting that the value $y = 0.171$ yields the exponents $\zeta_p$, 
which fully agree with their numerical values for all $p$, Fig.~\ref{fig7}. 
Computations show that $y = 0.185$ provides very accurate values for anomalous exponents of the Gledzer-Ohkitani-Yamada  
shell model~\cite{gledzer1973system,*ohkitani1989temporal}. Finally, $y = 1/3$ corresponds 
to the Kolmogorov scaling $\zeta_p = p/3$.

In conclusion, we showed that anomalous statistics of turbulence in shell models 
is determined exclusively by instanton dynamics. Instantons represent correlated extreme events 
reaching the smallest scale in finite time and having self-similar structure 
described by a single universal scaling exponent. 
We showed that the anomaly in the inertial range is determined by the process of instanton creation. 
We proposed a simplified model of instanton creation and computed the 
corresponding anomalous scaling exponents. 

\vspace{5mm}
This work was supported by CNPq under grant 477907/2011-3.

\bibliography{refs}
\bibliographystyle{apsrev4-1}

\end{document}